\RequirePackage{fix-cm}
\documentclass[twocolumn,epjc3]{svjour3}  
\smartqed  
\RequirePackage{graphicx}
\journalname{Eur. Phys. J. A}
\usepackage{amsmath,amssymb,bm}
\usepackage{graphics}
\usepackage{graphicx}
\usepackage{tabularx}
\usepackage{multirow}
\usepackage{psfrag}
\usepackage{comment}
\usepackage{pifont}
\usepackage{breqn}
\usepackage{url}
\usepackage{epsfig}
\usepackage{array}
\usepackage{float}

\usepackage{hyperref}
\usepackage[english]{babel}

\usepackage{epsfig}
\usepackage{cite}
\usepackage{color}
\usepackage{amssymb}

\numberwithin{equation}{section}
\numberwithin{equation}{subsection}

\renewcommand*{\theequation}{%
  \ifnum\value{subsection}=0 %
    \thesection
  \else
    \thesubsection
  \fi
  .\arabic{equation}%
}

\newcommand{\be}{\begin{equation}}
\newcommand{\ee}{\end{equation}}
\newcommand{\ba}{\begin{eqnarray}}
\newcommand{\ea}{\end{eqnarray}}

\newcommand{\pa}{\ensuremath{\tilde{a}_{_1}}}
\newcommand{\pb}{\ensuremath{\tilde{a}_{_2}}}
\newcommand{\psigma}{\ensuremath{\tilde{a}_{_3}}}

\begin{document}

\title{Negative moments as the signature of the radial density
at small distances
}


\author{M.~Atoui, M.~Hoballah, M.~Lassaut
        and
       J. Van de Wiele 
}


\thankstext{Corresponding author}{e-mail: mariam.atoui@gmail.com}


\institute{Universit\'e Paris-Saclay, CNRS/IN2P3, IJCLab,
91405 Orsay, France}


\maketitle

\begin{abstract}
The present paper proposes a robust evaluation of any radial density at small distances using negative-order radial moments evaluated in momentum space. 
This evaluation provides a valuable insight into the behavior of a given radial density in the vicinity of $r=0$, 
and puts strong emphasis on the importance of measuring form factors 
at large squared four-momentum transfer, a domain essential for the determination of negative order moments.
A specific attention is paid to the regularization scheme directly affecting the numerical determination of the radial density's parametrization. 
The proposed method is applied to non-relativistic study cases of the nucleon electric ($G_{En}, G_{Ep}$), and proton magnetic $G_{Mp}$ form factors.
The validation is performed through comparison of the results of the approach to the analytically determined 
Maclaurin expansion - in the vicinity of $r=0$ - of the radial density function.
The method is also applied to the relativistic Dirac form factor $F_1$ of the proton. In such a non-trivial case, 
the Maclaurin development might not exist for the radial density, rendering 
the determination from the proposed method extremely important.
\end{abstract}

\section{Introduction}\label{sect1}
\label{intro}
The determination of radial moments $<r^\lambda>$, \textcolor{black}{$\lambda$ real}, of the nucleon charge density distributions $f(\bf{r})$ offers key information on the nuclear electromagnetic structure.  
Previous conducted research has focused on the study of even-valued moments $<r^{\lambda}>$, \textcolor{black}{$\lambda$ integer}, from elastic electron-nucleus scattering~\cite{Liu:2021ofe}. For example, the charge 
root-mean-square radius can be obtained from the second-order moment $<r^2>$. The fourth-order moment $<r^4>$ plays a significant role in investigating the nucleon structure and is strongly associated with the surface 
thickness of the nuclear density distributions and determines the diffraction radius of the heavy nuclei~\cite{Reinhard:2019ixi}.
Even-valued radial moments $<r^{2n}>$ are proportional to 
\textcolor{black}{$n$-th derivative of the
Electric Form Factor (EFF) with respect to squared four-momentum transfer
in the origin,}
$
\left. \frac{\mathrm{d}^n G_E(k^2)}{\mathrm{d}k^{2n}} \right\vert_{k^2=0}.
$
It is hence established that measuring form factors, form fitting $e-p$ scattering data with a functional form, at small values of $k^2$ gives precise information about radial densities. However, derivative terms may not match precisely the fitting function coefficients used in the form factor extraction because of experimental errors and the domain of $k^2$ considered in the analysis.\\
All moments are of interest as they carry complementary information on the charge distribution inside the proton. The behavior of the charge distribution at small distances, for example, is encoded in low- and negative- order moments $\lambda$. 
A novel method, hereafter referred to as Integral Method (IM), was introduced in \cite{hoballah}. It enables the extraction of the radial moments relying on a principal value regularization of an integral, in momentum space, of form factors. 
One advantage of this method is that it generalizes to any real-valued moment order $\lambda > -3$ contrary to the standard method that utilizes derivative forms of the form factor and accesses only even-valued moment orders $\lambda \geq 0$. An application of the IM has been done by performing an extraction, using real data, of the moments of the spatial density of both even and odd orders~\cite{atoui}. \\
In this work, we aim to show that 
the information that can be obtained on the radial density and provided by the knowledge of even-valued moment orders can be enriched 
by extending the search to negative-order moments. 
We will demonstrate that for some negative values, the moments give significant information on the behavior of radial density.
The importance of this approach is demonstrated on the basis of generic and specific examples, and its validity is further discussed.\\
In the next section, we introduce the regularization scheme of the spatial moments
for all the negative orders $\lambda$ for both non relativistic and relativistic cases.
We stress the fact that one of the advantages of the proposed method is that the analytical form of these moments is conserved after regularization.
In a further section, we describe the regularization in momentum space.
Finally, we explain in detail how to extract numerically the radial density at small distances through a set of examples.

\section{Spatial moments}\label{sect2}
\subsection{General formalism}\label{sect2sub1}
Let $f_D(r)$ be a pure radial function, assumed to be integrable and fastly decreasing~\cite{schwartz} in the $D$-dimensional ($D=\{2,3\}$) space. The integral,
\be\label{sect2sub1eq01}
I=\omega_D \, \int_0 ^{\infty} f_D(r ) \, r^{D-1} \ dr
\ee
is finite by definition, where the factor $\omega_D$ denotes the $D$-dimensional solid angle, given by :
 \be \label{sect2sub1eq02}
\omega_D=\frac{2 \pi^{D/2}}{\Gamma \left(\frac{D}{2} \right)}
\hspace{10mm}   \omega_2 = 2\pi , 
\hspace{10mm}   \omega_3 = 4\pi .
\ee
The Fourier transform $\tilde{f}_D(k )$  of $f_D(r)$ exists \textcolor{black}{for every real value of $k$}. The moments $(r^\lambda,f_D)$ of the radial function $f_D$ are defined by
\be
\label{sect2sub1eq03}
( r^\lambda,f_D )= \omega_D \int_0 ^\infty f_D(r) \, r^{\lambda + D -1} \ dr .
\ee

\begin{figure}[h]
   \begin{center}
      \includegraphics[width=1.0\columnwidth]{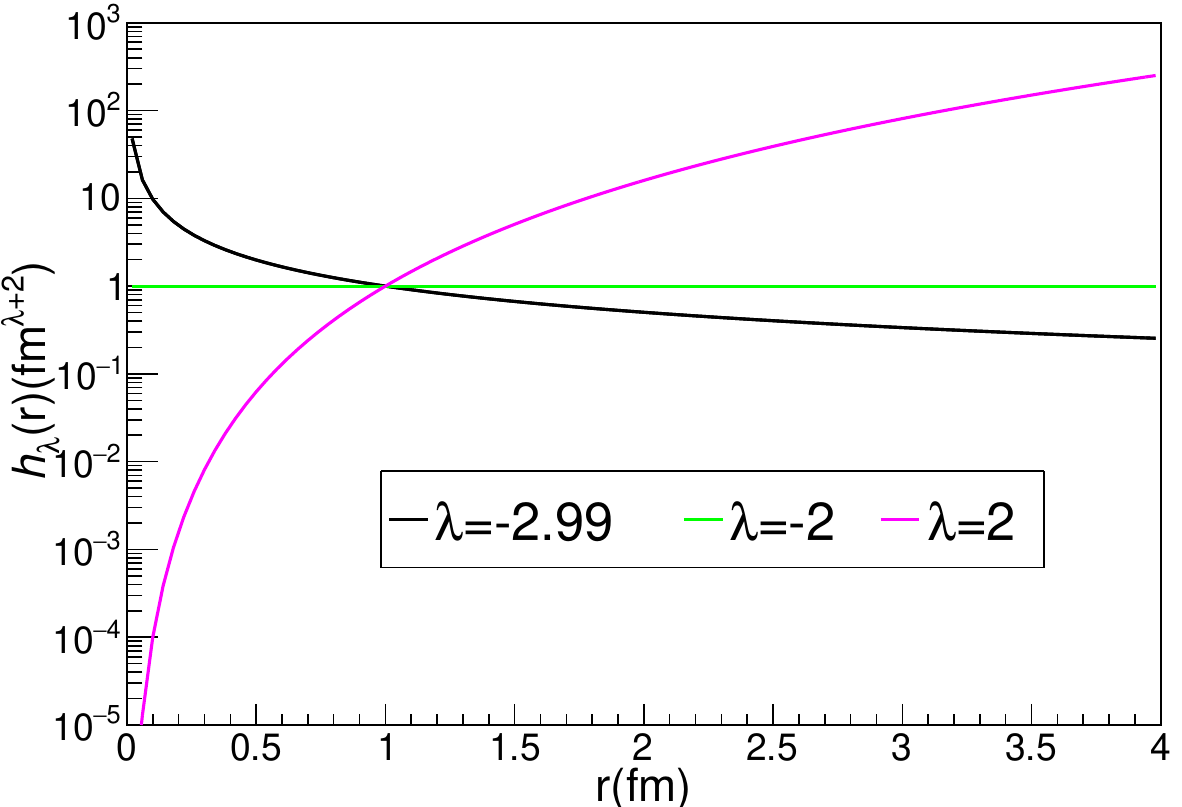}
   \end{center}
\caption{The term $h_\lambda(r)= r^{\lambda+2} $ for some considered positive and negative values of $\lambda$. For negative values of $\lambda$, 
the contribution to the radial function is higher at small distances $r$ meaning that moments with such particular order have more sensitivity to the radial density at such distances.}\label{fig:1}
\end{figure}

Examining Eq.~\ref{sect2sub1eq03}, it is observed that the term $h_\lambda(r)=r^{\lambda+D-1}$ probes the radial density $f_D(r)$. 
Figure~\ref{fig:1} shows the plot of $h_\lambda(r)$ for $D=3$ and for different positive and negative values of the moment order $\lambda$. 
It is shown that moments with negative orders have a particular sensitivity to radial density function at small distances as they emphasize the contribution of $h_\lambda(r)$. 
Although the moment with order $\lambda=2$ gives a hint of the charge radius, it can be seen that it can give no specific insight into how the radial density behaves near the center of the nucleon.
As a global feature, the higher contributions to the radial density originating from moment orders of $\lambda \ge 2$ are essentially given at large distances $r$. 

Analytically, information concerning the radial density $f_{D}(r)$ can be obtained thanks to the calculation, when existing, of the $D$-dimensional inverse Fourier transform of $ \tilde{f}_D(k )$ using 
\be\label{sect2sub1eq04}
f_D(r)= \frac{\omega_D}{(2\, \pi)^D}  \int_0^\infty  \tilde{f}_D(k) \ k^{D -1} \ \mathcal{J}_D(k\, r) \ dk ,
\ee
where
\ba\label{sect2sub1eq05}
\mathcal{J}_D(k\, r)
=      \begin{cases}
        J_0(k \,r) &  \text{for} \; D=2  \\
        j_0(k \,r) & \text{for} \; D=3 
     \end{cases},
\ea
where $J_0(k \,r)$ is the $0$-th order simple Bessel function and $j_0(k \,r)$ is the $0$-th order spherical Bessel function~\cite{Abra}.

Generally, the form of the Maclaurin expansion $f_D(r)$ of the radial density $f_{D}(r)$ in the vicinity of $r=0$ can be written as 

\be\label{sect2sub1eq05p5}
f_D(r)= \frac{1}{r^\beta} (\sum_{j=0}^{m}g_j r^j + \mathcal{O}(r^{m+1})), \, \,\,\,\,\, \,\,\,\,  m \ge 0,
\ee
where $\beta$ is defined such that
\be\label{sect2sub1eq06}
\lim_{\, r \to 0} r^\beta \  f_D(r) = g_0.
\ee
Our goal is to determine $f_D(r)$ in Eq.~\ref{sect2sub1eq05p5}. This reduces down to the determination of $\beta$ and $g_j$.
The integral of Eq.~\ref{sect2sub1eq01} is finite when $ \beta < D $. Moreover, the spatial moments expressed in the integral of Eq.~\ref{sect2sub1eq03}
are convergent when $ \lambda > -D + \beta$. Therefore, there exists a critical value of $\lambda $ which we will denote $\lambda_c \equiv -D + \beta$
for which the spatial moments are convergent, in particular when $ \lambda > \lambda_c$.\\
As a first step, we will show how to recover $ \lambda_c$ - and consequently $\beta$ -
directly from the measured form factor $ \tilde{f}_D(k)$. 
Secondly, we will work out the development of the radial function in the vicinity of $r=0$ proving that we can obtain precisely all the relevant parameters $g_j$ from the evaluation of the negative order moments in momentum space.\\

\subsection{Regularizing the divergence}\label{sect2sub2}
The integral shown in Eq.~\ref{sect2sub1eq03} exists for all  $\lambda$'s situated in the interval 
 $\lambda \in ]\beta-D, +\infty[$ namely when $\lambda > \lambda_c $ and hence requires no regularization in this interval.
For values of $\lambda \leq  \lambda_c$, 
the integral of Eq.~\ref{sect2sub1eq03} has to be taken in the sense of distributions according to Ref.\cite{Gue62}.
We define
\be\label{sect2sub2eq02}
g(r)=f_D(r) \, r^{\beta},
\ee
and start with the case where $g(r)$
 is of fastly decreasing~\cite{schwartz} at infinity and is $C^{\infty}$ at zero. 
 Following Eq.~\ref{sect2sub1eq05p5}, for each value of $ m$, $g(r)$ is expanded, in the vicinity of $r=0$, as:
 \be\label{sect2sub2eq03} 
 g(r) = \sum_{j=0}^ {m} g_j r^j + \mathcal{O}(r^{m+1}), \, \,\,\,\,\, \,\,\,\,  m \ge 0.
\ee
Equation~\ref{sect2sub1eq03} now reads 
\begin{eqnarray}\label{sect2sub2eq06}
  (r^{\lambda},f_D)
   & = & \omega_D \  
  \int_0^{\infty}  dr \ 
  r^{\lambda-\lambda_c-1} g(r)  .
  \end{eqnarray}
For  $\lambda  >\lambda_c -m - 1 $, we have :
\begin{align}\label{sect2sub2eq04}    
 \frac{(r^{\lambda},f_D)}{\omega_D}
& = \hspace{1mm} \int_0^{1}  dr \ r^{\lambda-\lambda_c-1} 
\Big( \  g(r)-\sum_{j=0}^{m}  g_{j}\  r^j \ \Big) 
\nonumber\\
& \hspace{-8mm}
+\int_{1}^{\infty}  dr \ r^{\lambda-\lambda_c-1}  g (r)
 +  \sum_{j=0}^{m}   \frac{ g_{j}}{\lambda-(\lambda_c-j)}.
\end{align}
The first integral  converges for $ \lambda > \lambda_c -m - 1  $
and \textcolor{black}{given that $g(r)$ is fastly decreasing~\cite{schwartz},} the second integral converges for all values of $\lambda$.
So we deduce that Eq.~\ref{sect2sub2eq04} has a meromorphic continuation to  $\lambda-\lambda_c+m + 1  >0$ with simple poles
of residue $g_j$ at  $\lambda-\lambda_c+j=0$ with $ j = 0,1, 2, . . . , m$ and no other singularities. 
In the vicinity of a pole, for instance $\lambda = \lambda_c - j_0 $,
$ 0 \leq j_0 \leq m$,
\be\label{sect2sub2eq05}    
 \frac{(r^{\lambda},f_D)}{\omega_D}
\approx  
\frac{ g_{j_0}}{\lambda-(\lambda_c-j_0)}.
\ee
The moments as expressed in 
Eq.~\ref{sect2sub1eq03} are nothing 
but the Mellin transform of the function $f_D(r)$.
Much work has been done on the Mellin transform~\cite{Mellin,Wong}. Particularly, 
the analytical extension with complex arguments has been investigated
in the case where the development of the function $f_D(r)$ in the vicinity of 
zero admits logarithmic terms (not $C^{\infty}$ anymore).
This concerns the Bessel function $K_n$, of integer index $n$, which is fastly decreasing. 
The Bessel functions are encountered here as inverse Fourier transforms of 2-dimensional monopoles or dipoles. 
They have the property to include logarithmic terms in the development of the 
function $g$ in the vicinity of $r=0$, it follows that in such a case, Eq.~\ref{sect2sub2eq03} reads
\begin{eqnarray}\label{sect2sub2eq07}
 g(r) & \simeq & \sum_{j=0}^m  \sum_{\ell=0}^n g_{j,\ell} \ \ln^{\ell}(r)   \ r^j + \ldots  .
\end{eqnarray} 
The integral in Eq.~\ref{sect2sub2eq06} remains convergent for $ \lambda>\lambda_c$. 
For $\lambda \leq \lambda_c$ the integral has to be regularized.
Taking into account the results concerning Mellin transform we have 
\begin{align}\label{sect2sub2eq08}
\frac{(r^{\lambda},f_D)}{\omega_D}
&  =       \int_0^{1}  dr \ 
r^{\lambda-\lambda_c-1} 
\Big( g(r)-\sum_{j=0}^m \sum_{\ell=0}^n g_{j,\ell} \ln^{\ell}(r)  \  r^j 
\Big) 
\nonumber\\
& \hspace{2mm}
+\int_{1}^{\infty}  dr \ r^{\lambda-\lambda_c-1}  g (r) 
\nonumber\\
&\hspace{2mm}
+  \sum_{j=0}^m \sum_{\ell=0}^n   
\frac{ (-)^{\ell} \ell! g_{j,\ell}}{(\lambda-(\lambda_c-j))^{\ell+1}} .
\end{align}
Equation~\ref{sect2sub2eq08} defines an analytical extension of the moments in the domain 
$ \lambda > \lambda_c-m -1   $, where the first integral is convergent.
It shows that the presence of a logarithmic term in the expansion is the signature of double poles for
 the moments.
 The presence of logarithmic squared terms is the signature of triple poles for the moments etc.
The moments defined by Eqs.~\ref{sect2sub2eq08} correspond to meromorphic functions of $\lambda$
in the domain $ \lambda > \lambda_c-m -1   $,
having poles at $\lambda=\lambda_c-j, \, j=0,1,2,\cdots,m$,
of multiplicity at most $n+1$. \\
In the vicinity of a pole, for instance $\lambda = \lambda_c - j_0 $,
$ 0 \leq j_0 \leq m$,
\begin{align}\label{sect2sub2eq09}    
 \frac{(r^{\lambda},f_D)}{\omega_D}
\approx 
\sum_{\ell=0}^n   
\frac{ (-)^{\ell} \ell! g_{j_0,\ell}}{(\lambda-(\lambda_c-j_0))^{\ell+1}}.
\end{align}
In summary, the moments are expressed in terms of the coefficients $g_{j_0}$ (Eq.~\ref{sect2sub2eq05}) and $g_{j_0,\ell}$ (Eq.~\ref{sect2sub2eq09}). The determination of the moments, and therefore these coefficients will be an essential step to reconstruct the radial densities at small distances. We will do so starting from the expression of moments in terms of the measured Fourier transform $\tilde f_D(k)$ as we will explain in the forthcoming sections.
\section{Radial moments in momentum space}\label{sect3}
An alternative approach consists in expressing the moments of the radial density $f_D(r)$ in terms of its $D$-dimensional Fourier transform $f_D(k)$
\cite{hoballah}.
According to Eq.~14 of \cite{hoballah}, the moments of the spatial density are given by
\be\label{sect3eq01}
(r^{\lambda},f_D) =
 \mathcal{N}_{\lambda ; D} \int_{0} ^{\infty} dk \, 
 { \left\{ \frac{ \tilde{f_D}(k) } {k^{\, \lambda +1 }} \right\} } ,
\ee
where $\mathcal{N}_{\lambda; D}$ is the normalization coefficient
\begin{equation}\label{sect3eq02}
\mathcal{N}_{\lambda;D} =2^{\lambda+1} \, \frac{\Gamma(\frac{\lambda+D}{2})}{\Gamma(-\frac{\lambda}{2}) \Gamma \left( \frac{D}{2} \right)},
\end{equation}
and the integral in Eq.~\eqref{sect3eq01} is taken in the sense of distributions, {\it i.e.} the principal value of the integral defined from the regularization of the diverging integrand at infinite-momentum. For negative values of $\lambda$, no divergence is encountered when $k$ is close to zero. \\
For the sake of clarity, henceforth we will adopt
the convention $\{ h(k) \}_a$  referring to the regularization 
of the function $h$ in the vicinity of $k=a$.
In contrast with the reference \cite{hoballah}, here we are concerned
with negative values of $\lambda$ and consequently  
with the regularization of \ref{sect3eq01} at infinity.\\
\textcolor{black}{We assume that the  physical distribution of moments is such that the class of functions $\tilde{f_D}$ satisfies the property that there exists a value of $k$ say $K$}
for which we have 
\begin{equation}\label{sect3eq03}
 \tilde{f_D}(k)
= \sum_{j=0}^{\infty} \frac{[\tilde{f_D}]_{-j}}{ \, k^{p+j}},
\qquad\quad p >1,
\hspace{10mm}
\end{equation}
\textcolor{black}{the series being convergent for every $k \geq K$.} Consequently
\begin{equation}\label{sect3eq04}
\left\{ \frac{ \tilde{f_D}(k) }{k^{\, \lambda +1 }} \right\}_{\infty} \equiv \frac{1}{k^{\lambda +1 }} \ 
\left( \tilde{f_D}(k) - 
\sum_{j=0}^{n} \frac{[\tilde{f_D}]_{-j}}{ \, k^{p+j}}  
\right),
\end{equation}
where $n$ is the integer such that $-p-1-\lambda < n \leq -p - \lambda$ and
positive 
when $\lambda+p$ is  strictly negative. In other words, $n$ is positive 
 and is the integer part of the strictly positive number   $-p  -\lambda$.\\
At the present stage, we have
\begin{align}\label{sect3eq05}
 (r^{\lambda},f_D) &=
\omega_D
 \int_0^{\infty} dr  \left\{ r^{\lambda+D-1-\beta} g(r) \right\}_0  
 \nonumber\\[0mm]
& 
= 
2^{\lambda+1} \,\frac{ \Gamma \left(\frac{\lambda+D}{2} \right)}
      {\Gamma \left( -\frac{\lambda}{2} \right)
         \Gamma \left( \frac{D}{2}\right) }
\int_{0} ^{\infty} dk \, { \left\{ \frac{ \tilde{f_D}(k) } {k^{\, \lambda +1 }} \right\} }_{\infty},
\end{align}
where both integrals in Eq.~\ref{sect3eq05} are taken in 
the sense of distributions. We show, in~\ref{demo_reg_mom}, that spatial moments can be expressed as:
\begin{align}\label{sect3eq16}
 (r^{\lambda},f_D) 
&=
 2^{\lambda+1}
 \frac{\Gamma \left(\frac{\lambda+D}{2} \right)}
      {\Gamma \left( -\frac{\lambda}{2} \right) 
        \Gamma \left( \frac{D}{2}\right)}
\nonumber \\
& \hspace{-8mm}
  \times
\Bigg[
  \int_0^{K} \! \! \! dk 
  \,  \frac{\tilde{f_D}(k)}{k^{\lambda+1}} 
  + \sum_{j=0}^{\infty} \frac{[\tilde{f_D}]_{-j}}
                       { \, (  \lambda+p+j) \, K^{\lambda+p+j}}
\Bigg].
\end{align}

\section{Physical Application}\label{sect4}

\subsection{A general example in $D=3$ }\label{sect4sub1}
\begin{figure}[b]
  \begin{center}
     \includegraphics[width=1.0\columnwidth]{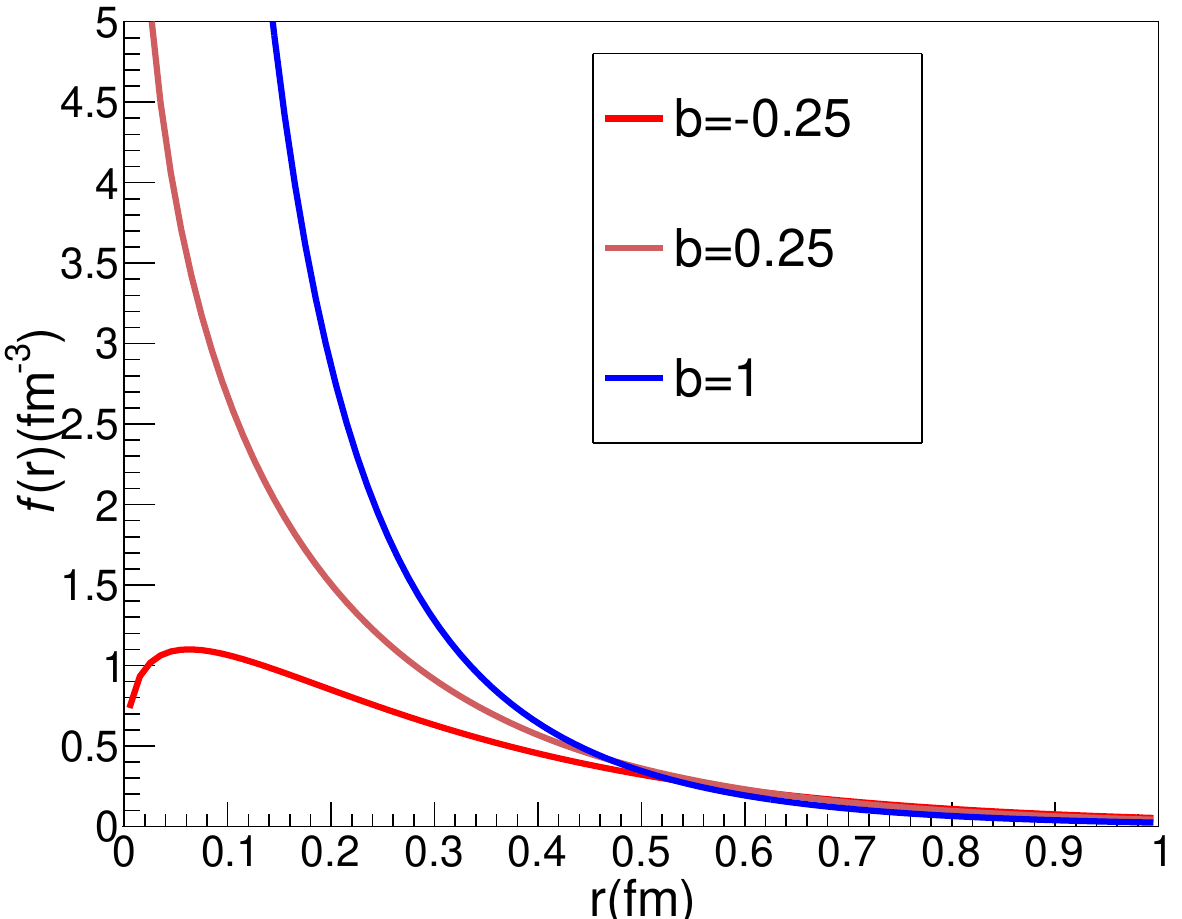}
  \end{center}
\caption{The behavior of $f(r)$, Eq.~\ref{sect4sub1eq01}, for different values of $b$ and for $a=4\, fm^{-1}$.}\label{fig:fder}
\end{figure}

\begin{figure}[H]
   \begin{center}
      \includegraphics[width=1.0\columnwidth]{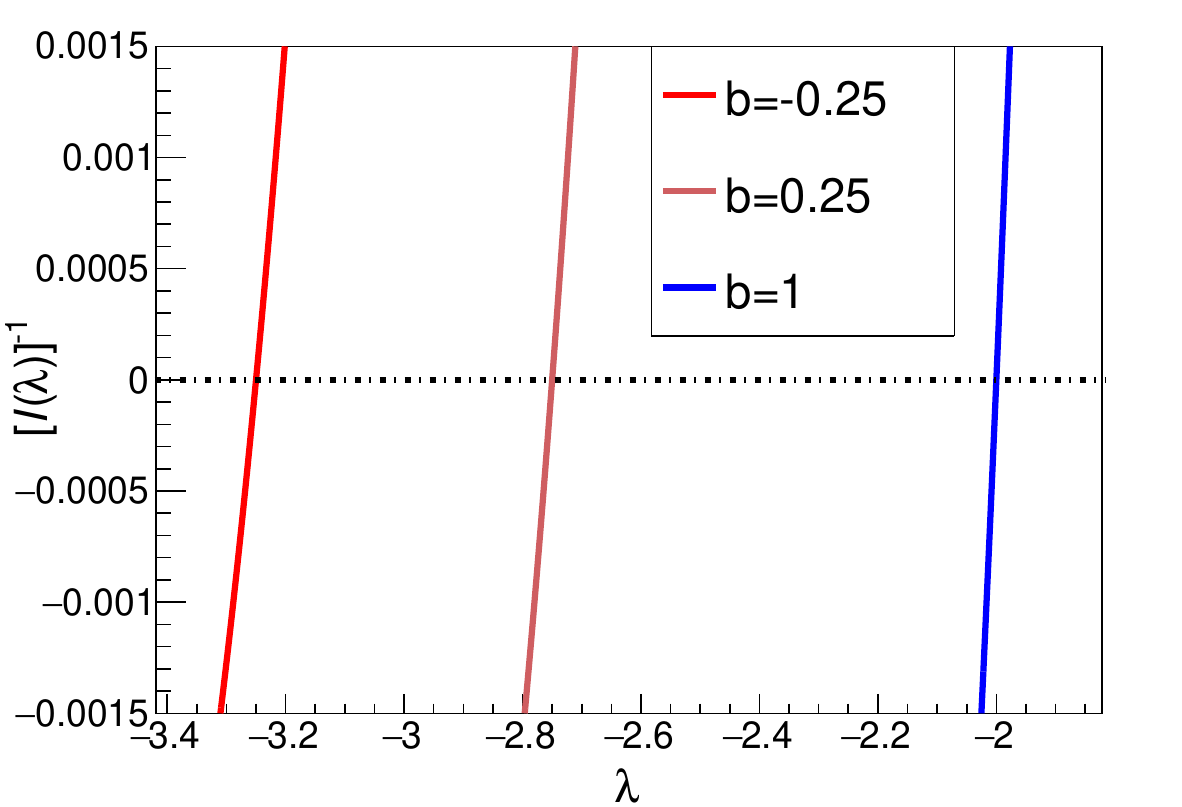}
   \end{center}
\caption{
$\left[\mathcal{I}(\lambda )\right]^{-1}$ for
$i_{min}=-12$ and for different values of $b$.
}\label{fig:2}
\end{figure}

Before investigating the experimentally measured form factors, we illustrate the method, 
for the sake of pedagogy, with a simpler example. As a matter of fact, radial densities associated with the nucleon form factor 
behave like either $\propto e^{-ar} $ or 
$\propto e^{-ar}/r $ for $r\to0$, 
so either $\beta = 0 $ or $\beta = 1$.
 Let us consider the function 
\begin{align}\label{sect4sub1eq01}
f(r) = \frac{a^{3-b}}{4 \, \pi \ \Gamma(3-b)} \frac{e^{-a \, r}}{r^b}
\hspace{6mm}
a>0,
\hspace{6mm}
b<3,
\end{align}
satisfying 
\begin{align}\label{sect4sub1eq02}
4 \, \pi
\int_0^\infty f(r) \, r^2 \ dr = 1.
\end{align}
Figure~\ref{fig:fder} shows the behavior of $f(r)$ for different values of $b$ and for $a=4\, fm^{-1}$.
The analytical expressions of the $g_j'$s are known from the Maclaurin
expansion of \ref{sect4sub1eq01}. The measured form factor which is the Fourier transform of $f(r)$ is given by
\begin{align}\label{sect4sub1eq03}
\tilde{f}(k)
=
\frac{a^{3-b}}{b-2} \
\frac{\left( a^2 + k^2 \right)^{-1+b/2} \sin \big( (b-2) \mathrm{arctan}(k/a )  \big)}{k},
\end{align}
and 
its asymptotic development reads:
\begin{align}\label{sect4sub1eq04}
\tilde{f}_{\rm asymp}(k)
&= 
\bigg\{ \sum_{i=i_{\rm min}}^{-3} c_i \ k^{i+b} \ \bigg\}  
\cos \left(\frac{b \pi}{2} \right) 
\nonumber\\
& +
\bigg\{ \sum_{i=i_{\rm min}}^{-3} s_i \ k^{i+b} \ \bigg\}  
\sin \left(\frac{b \pi}{2} \right),
\end{align}
where the coefficients $c_{i} = 0$ when $i$ is odd,
and $s_{i} = 0$ when $i$ is even.\\
The spatial moments in eq. \ref{sect3eq01} can then be written as
\ba
\mathcal{I}(\lambda) &= &\mathcal{N}_{\lambda;D} \
\Bigg\{ \ \ \int_0 ^{K}  \frac{\tilde{f}_D (k)}{k^{ \lambda + 1}}  \ dk 
\nonumber\\
& + &
\bigg\{ 
\sum_{i=i_{min}}^{-3} c_i \ \frac{K^{b+i - \lambda}}{\lambda - b -i} 
     \ \bigg\}  
\cos \left(\frac{b \pi}{2} \right)
\nonumber \\[1mm]
& 
+&
\bigg\{ \sum_{i=i_{min}}^{-3} s_i \ \frac{K^{b+i - \lambda}}{\lambda - b -i} 
\ \bigg\}  
\sin \left(\frac{b \pi}{2} \right)
\ \ \Bigg\},
\ea
and will be evaluated numerically. We then calculate $ \left[\mathcal{I}(\lambda)\right]^{-1}$ for a 
series of values of  $\lambda $ inferior to $-3/2$ for instance. Thanks to this scanning method,
 $\lambda_c $ is obtained as the solution of the equation 
\begin{align}\label{sect4sub1eq07}
\left[\mathcal{I}(\lambda_c)\right]^{-1} = 0
\end{align}
implying that
\begin{align}\label{sect4sub1eq08}
\beta = \lambda_c + D = \lambda_c + 3.
\end{align}
Figure~\ref{fig:2} shows the distribution of $\left[\mathcal{I}(\lambda_c)\right]^{-1}$ for different values of the parameter $b$ and for  $a=4\, fm^{-1}$.\\
The calculation of  $g_0$, then the other $g_i'$s is made as follows:
one chooses $2 N$ values of $ \lambda $ around  $\lambda_c$  namely  $N$ 
values
$ \lambda_1, \cdots, \lambda_{N} $ such that $ \lambda_i < \lambda_c $
and $N$ values
$ \lambda_{N+1}, \cdots, \lambda_{2N} $ such that $ \lambda_i > \lambda_c $.
For each values one calculates (cf. Eq.\ref{sect2sub2eq05}):
\begin{align}\label{sect4sub1eq09}
1) \hspace{1mm}
\eta_i \equiv \lambda_i - \lambda_c,
\hspace{15mm}
2) \hspace{1mm}
y_i \equiv \frac{\eta_i}{4 \, \pi } \ \mathcal{I}(\lambda_i).
\end{align}
Then construct the function  $(x_i, y_i) $ where $x_i \equiv \lambda_i$. One constructs the associated Lagrange interpolation polynomial of degree $5$ for instance
$ y_i = \ell_5(x_i) $ then obtains $g_0$ thanks to 
\begin{align}\label{sect4sub1eq10}
g_0 = \ell_5(\lambda_c).
\end{align}
This procedure is generalized to give 
\begin{align}\label{sect4sub1eq11}
g_1 = \ell_5(\lambda_c - 1), \hspace{5mm}  g_2 = \ell_5(\lambda_c - 2), \hspace{5mm}
\cdots .
\end{align}
The numerical results are presented in table 
\ref{tab:2}. It is important to note that those results are in full agreement with the analytical results evaluated using the Maclaurin
expansion of \ref{sect4sub1eq01}; the latter not shown for redundancy.


\begin{table}[H]
\begin{center}
\begin{tabular}{|c|c|c|c|c|} \hline
 $ \lambda_c$ & $  \beta  $ & $g_0    $ & $  g_1    $  & $ g_2    $  \\
  & $   $ & $(fm^{-3+\beta})   $ & $  (fm^{-4+\beta})   $  & $  (fm^{-5+\beta})   $  \\
\hline
 $ -3.50   $ & $ -0.50    $ & $ 3.0649$ & $ -12.2598 $  & $ 24.5196 $ \\
 $ -3.25   $ & $ -0.25    $ & $ 2.8253$ & $ -11.3014 $  & $ 22.6028 $ \\
 $ -3.00   $ & $  0.00    $ & $ 2.5465$ & $ -10.1859 $  & $ 20.3718 $ \\
 $ -2.75   $ & $  0.25    $ & $ 2.2391$ & $  -8.9564 $  & $ 17.9127 $ \\ 
 $ -2.50   $ & $  0.50    $ & $ 1.9156$ & $  -7.6624 $  & $ 15.3248 $ \\
 $ -2.00   $ & $  1.00    $ & $ 1.2732$ & $  -5.0929 $  & $ 10.1859 $ \\ 
\hline 
\end{tabular}
\caption{The numerical results of $g_0,\; g_1$, and $g_2$ for different values of $\lambda_c$, 
and $\beta$; $ a $ is taken to be $= 4 $ fm $^{-1}$. These results are in full agreement with 
those obtained via analytical evaluation of the Maclaurin expansion of \ref{sect4sub1eq01}.}
\label{tab:2}
\end{center}
\end{table}

\subsection{Electric and magnetic proton form factors}\label{sect4sub2}
\begin{figure*}[t]
  \includegraphics[width=0.66666666\columnwidth]{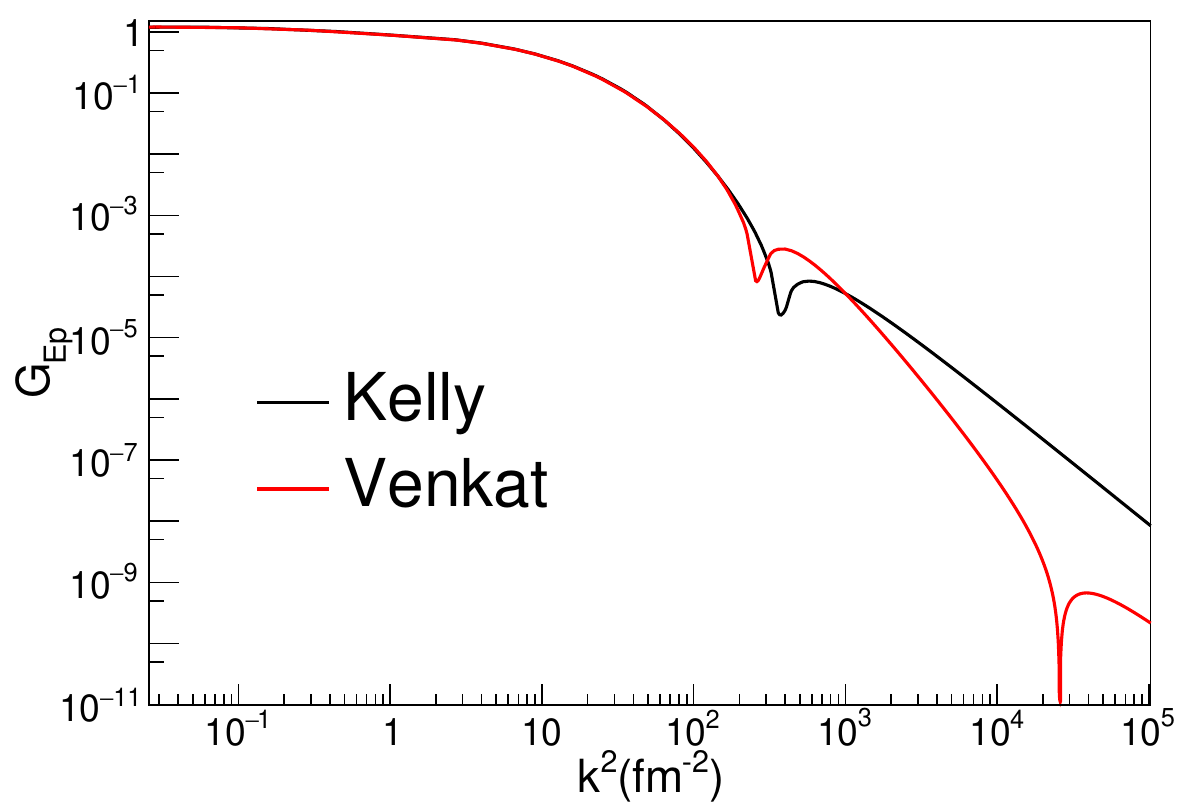}
  \includegraphics[width=0.66666666\columnwidth]{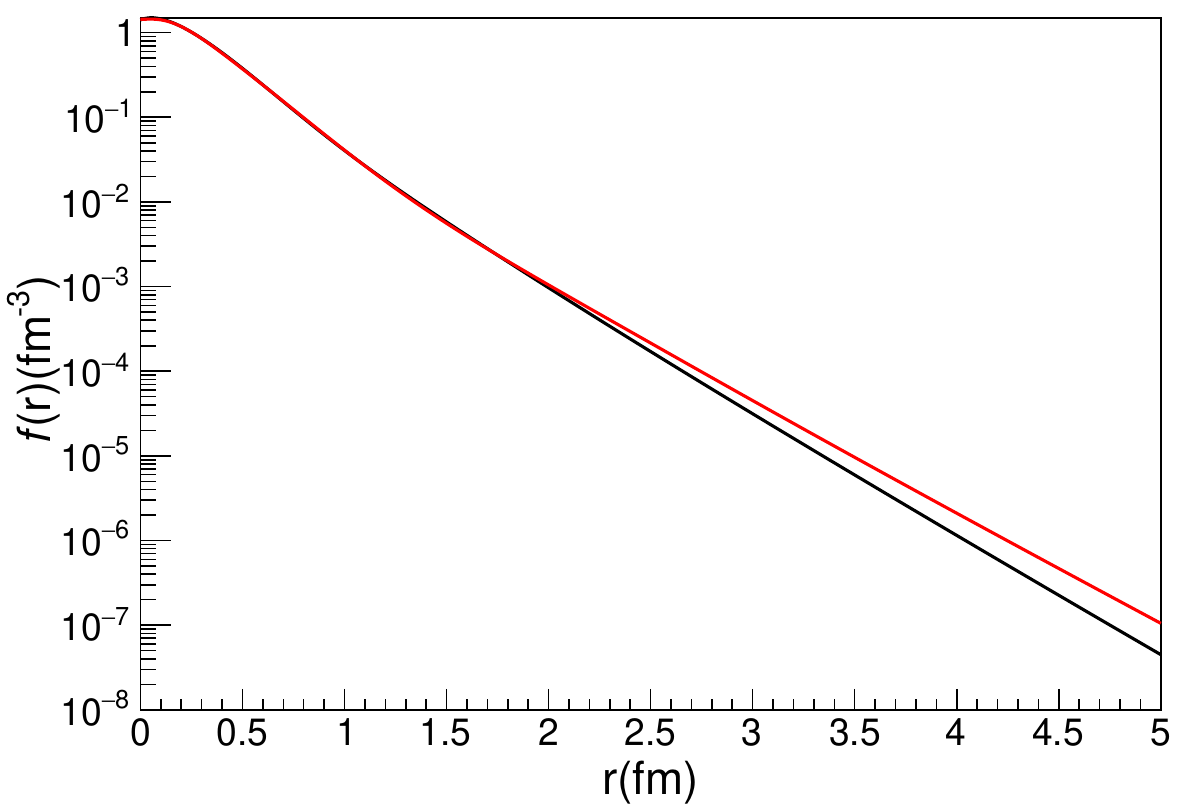}
  \includegraphics[width=0.66666666\columnwidth]{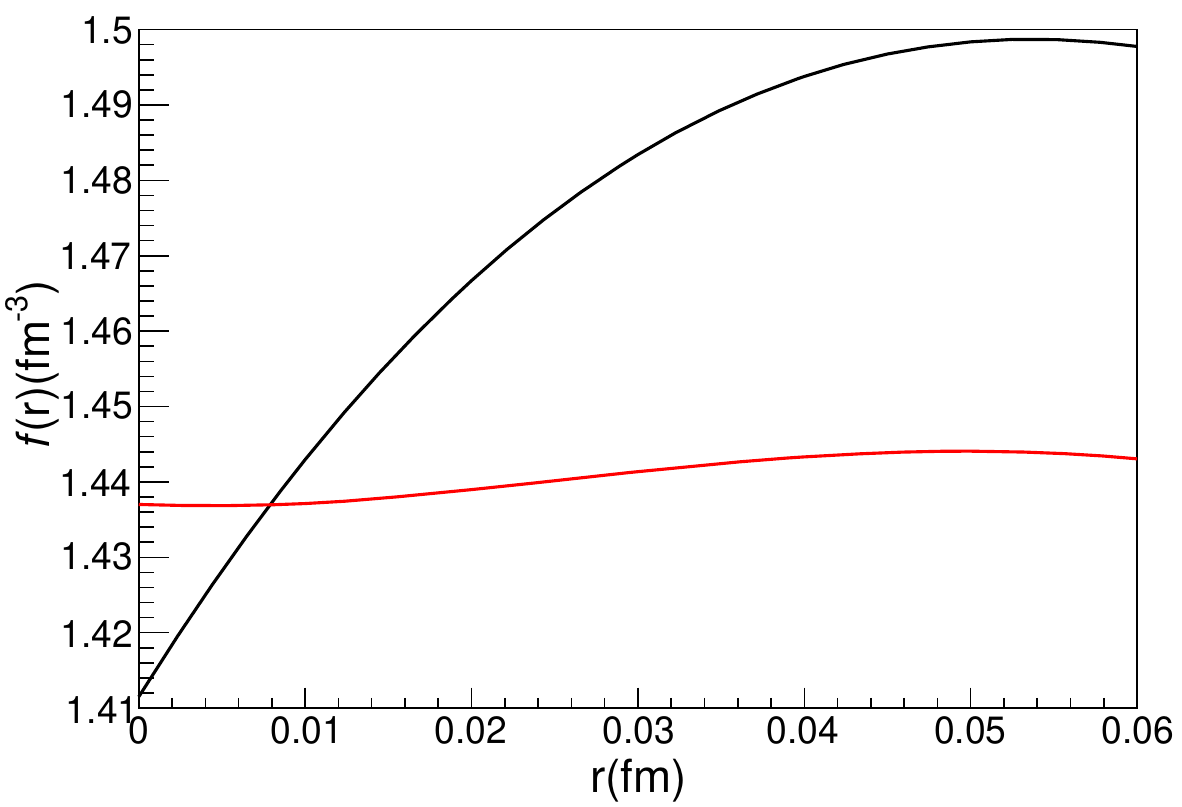}
      \caption{(Left) Absolute value of the electric form factor of proton
      from Ref.\cite{kellyEM} and \cite{venkat} drawn in natural units of fm$^{-2}$; 1GeV$^2 = 1/(\hbar c)^{2}$fm$^{-2}$.
      (Middle) Radial densities corresponding to the two considered parametrizations of $G_{Ep}$. 
      (Right) Radial densities at small distances (for $r \ll 1$ fm).}\label{fig:34}
\end{figure*}

We will apply our method to both electric  $G_{Ep}$ and magnetic $G_{Mp}$ proton form factors starting from an analytical form $\tilde f(k)$ corresponding to the ratio of two polynomials.
This form reproduces the experimentally measured electric  $G_{Ep}$ and magnetic $G_{Mp}$ proton form factors\cite{higginbotham}. Moreover, the inverse Fourier transform $f(r)$ can be calculated analytically and this allows to test the precision of the method.
Two parametrizations discussed in Ref.\cite{kellyEM} and \cite{venkat} will be used for the proton form factors:
\begin{eqnarray}
\tilde{f}(k) \quad &=& \quad \eta_p \quad
\frac{1 + \sum_{i=1}^{n-2} \tilde{a}_i \, k^{\, 2\, i}}
     {1 + \sum_{i =1}^{n} \tilde{b}_{i} \, k^{\, 2\, i}},
\nonumber\\[2mm]
\eta_p \quad &=& \quad
\left\{ \quad \begin{array}{lr}
           1      & \quad\tilde{f}(k)=G_{Ep}(k^2) \\
	   \mu_p   & \quad\tilde{f}(k)=G_{Mp}(k^2)
	 \end{array}
\right. .
\label{sect4sub2eq01}
\end{eqnarray}
All the  coefficients $ \tilde{a}_i $ and $ \tilde{b}_i$ are real 
and all the coefficients $ \tilde{b}_i $ are positive.
The $ n $ values considered here are  $n=3$ (Ref.\cite{kellyEM})
 and $n=5$ (Ref.\cite{venkat}). They form factor parameterizations are shown in the left panel of 
Fig.~\ref{fig:34}. 
At large values of the squared four-momentum transfer $k^2$, they are different and we can expect
to find large differences in their inverse Fourier transform when $r\to 0$.\\

The series expansion of  $\tilde{f}(k)$ when $k \to \infty$ is given by
\begin{align}\label{sect4sub2eq02}
\tilde{f}(k)
\approx \sum_{i=i_{\rm min}}^{i=-4} 
c_{i} \, k^{i},
\hspace{10mm} i \hspace{2mm} \text{is even.}
\end{align}
The function $ \mathcal{I}(\lambda) $ then reads
\begin{align}\label{sect4sub2eq04}
 \mathcal{I}(\lambda) \approx
 \mathcal{N}_{\lambda;3} \
\Bigg\{ 
\int_0 ^{K} 
\frac{\tilde{f} (k)}{k^{ \lambda + 1}}
\ dk
+ 
\sum_{i=i_{\rm min}}^{-4} c_i \ \frac{K^{i - \lambda}}{\lambda  -i}   
 \Bigg\}     .
\end{align}
Using the Partial Fraction Decomposition of $ k \, \tilde{f}(k)$,
the radial density is known to be
\begin{align}\label{sect4sub2eq05}
f(r) =
\frac{1}{2 \, \pi } \frac{1}{r}
 \sum_{j=1}^{n}     e^{- k_{j {\mathrm{I}}} \, r }
 \
\Big[ 
&
   A_{jR} \, \cos(k_{j {\mathrm{R}}} \, r )
\nonumber\\[-3mm]
& \hspace*{0.5cm} 
 -  A_{jI} \, \sin(k_{j {\mathrm{R}}} \, r )
 \ \Big] ,
\end{align}
and its Maclaurin development gives the analytical expression
of the $g_j'$s.
In table~\ref{tab:3} we report the values of $\beta$ and $g_i$'s  determined numerically 
using $\mathcal{I}(\lambda)$ in Eq.~\ref{sect4sub2eq04} and corresponding to the electric and magnetic form 
factors of the proton. We obtain the same exact values using the analytical expression of $f(r)$ form Eq.~\ref{sect4sub2eq05}.
\begin{table}[h]
\begin{center}
\begin{tabular}{| l | r   |  r  |  r  |} \hline
$ \ $    & \multicolumn{2}{c|}{$G_{Ep}$} & \multicolumn{1}{c|}{$G_{Mp}$}\\
 $ \ $    & \multicolumn{1}{c}{Ref.\cite{kellyEM} }& \multicolumn{1}{c|}{Ref.\cite{venkat}} & \multicolumn{1}{c|}{Ref.\cite{kellyEM} }\\
 \hline
 $ \beta$ & $0.0000  $  & $0.0000$    &   $0.0000 $  \\
 $ g_0 (fm^{-3}) $ & $1.4115  $ & $1.4370$     &  $2.7962 $   \\
 $ g_1 (fm^{-4}) $ & $3.5549  $  & $-0.1215$   &  $-5.9620 $  \\
 $ g_2 (fm^{-5}) $ & $-42.2277$   & $15.8100$  &  $-32.0702$  \\
\hline 
\end{tabular}
\caption{The numerical evaluation of $g_0,\; g_1$, and $g_2$ for the two considered parametrizations of the 
proton electric form factor $G_{Ep}$ and that of the proton magnetic form factor $G_{Mp}$. 
These results are in full agreement with those obtained via analytical evaluation using Eq.~\ref{sect4sub2eq05}.
The value of $ \beta = 0 $ corresponds to $\lambda_c = -3$.}\label{tab:3}
\end{center}
\end{table}
Figure~\ref{fig:34} shows the radial denstities - middle and right pannels - obtained when considering the parametrizations 
of Eq~\ref{sect4sub2eq01} for the electric form factor. On a global level, these two parametrizations agree, however, strongly disagreeing at low $r$.
\subsection{Neutron electric form factor}\label{sect4sub3}
\begin{figure}[h]
   \begin{center}
      \includegraphics[width=1.0\columnwidth]{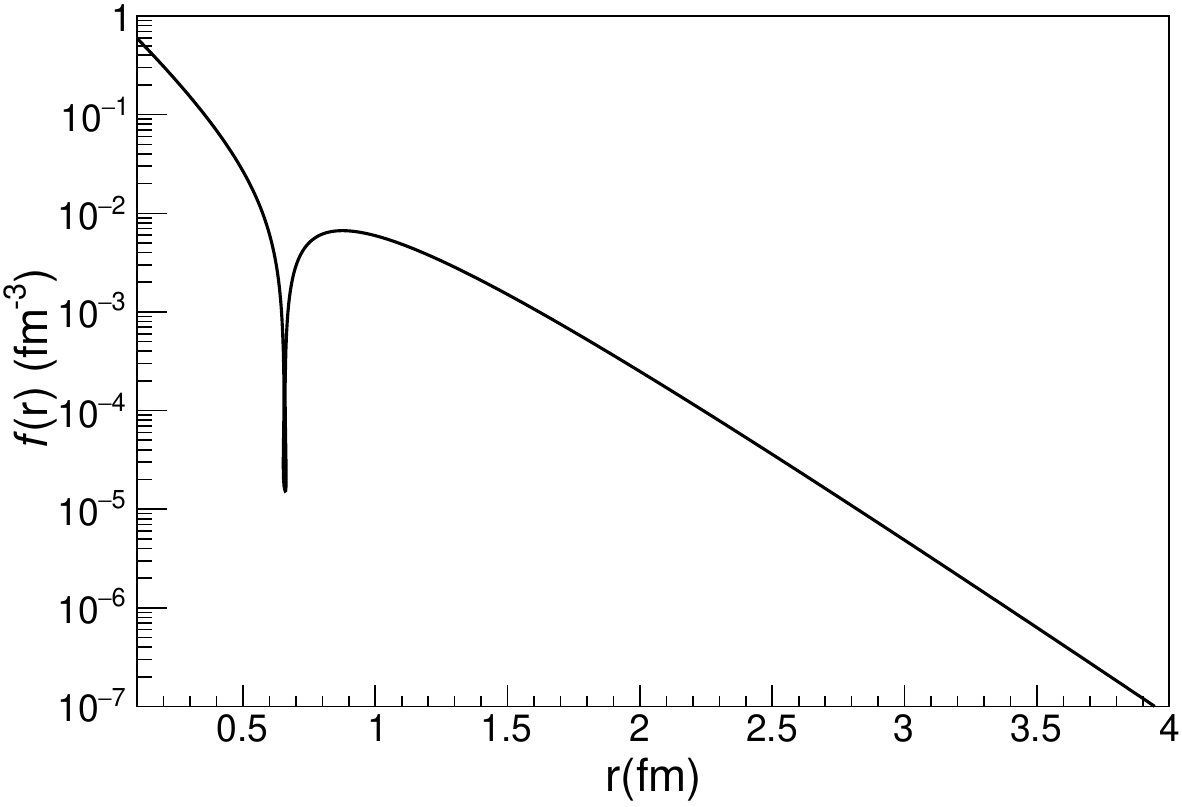}
   \end{center}
\caption{The absolute value of the radial density of equation \ref{sect4sub3eq03} associated with
 the neutron form factor $G_{En}$.}\label{fig:6}
\end{figure}

For the neutron electric form factor, we will consider the parametrization given in reference~\cite{kellyEM} which reads
\begin{align}\label{sect4sub3eq01}
\tilde{f}   (k) = \frac{\pa \, k^2 }{1 + \pb \, k^2} \ 
       \frac{ \psigma ^4 }{(\psigma ^2 + k^2 )^2} ,
\end{align}
with
\begin{align}\label{sect4sub3eq02}
\lim_{k \to \infty}\tilde{f} (k)
=
\frac{C_{-4}}{k^4} + \! \frac{C_{-6}}{k^6} + \cdots
+ \frac{C_{i_{\rm min}}}{k^{-i_{\rm min}}}.
\end{align} 
The radial density related to the neutron electric form factor is shown in Fig.~\ref{fig:6} and is given by
\begin{align}\label{sect4sub3eq03}
f(r) 
= &
\frac{\pa \, \psigma ^4}
     {4 \pi  \left(-1+\pb \, \psigma ^2\right)^2}
     \nonumber\\
     & \hspace{-8mm} \times
\left(
- \frac{e^{ - \frac{{\scriptstyle{r}}}{\sqrt{ \pb ^{\ }} }} } {r}
+ \frac{e^{-r  \psigma }}{r}
-\frac{\psigma}{2} \, e^{-r  \psigma }
+\frac{\pb \, \psigma ^3}{2}  \, e^{-r  \psigma } \right).
\end{align}
The numerical results that we have obtained are presented in table \ref{tab:5} and are in full agreement with the analytically evaluated ones.
\begin{table}[h]
\begin{center}
\begin{tabular}{| l | r  |} \hline
$ \ $  & \multicolumn{1}{c|}{$ \ \ G_{En}$   }\\
\hline
$ \beta $ & $ 0.0000$ \\
$ g_0 (fm^{-3})  $ & $ 1.1119$ \\
$ g_1 (fm^{-4}) $ & $-6.8150$ \\
$ g_2 (fm^{-5}) $ & $19.6290$ \\
\hline
\end{tabular}
\caption{The values of $g_0,\; g_1$, and $g_2$ for the considered parametrization of the neutron electric form factor $G_{En}$. These results are in full agreement with the analytical one obtained using Eq.~\ref{sect4sub3eq03}. The value of 
$ \beta = 0$ and is equivalent to $\lambda_c = -3$.}\label{tab:5}
\end{center}
\end{table}

\subsection{A relativistic interpretation of the nucleon size}\label{sect4sub4}

Moments of the charge density extracted 
using $G_{Ep}(k^2)$, $G_{Mp}(k^2)$ or $G_{En}(k^2)$ 
do not have a strict probabilistic interpretation because of the relativistic nature of the nucleon~\cite{Miller,Lor20}. 
For proton-like objects, where the intrinsic size is comparable to the associated Compton wavelength, 
a charge density distribution cannot be unambiguously defined~\cite{Jaf21}. 
To take into account relativistic effects, several prescriptions have been suggested. The infinite-momentum approach suggests that the Dirac form factor $F_1(k^2)$ might be more 
appropriate than $G_E(k^2)$~\cite{Miller}, however assimilating the proton to a disk.
Another approach proposing a new definition of the electromagnetic spatial densities also prefers the use of $F_1(k^2)$, quantitatively differing from the infinite-momentum approach~\cite{Pan22}. 
A recent work demonstrated that an unambiguous relativistic correction to the conventional Sachs distributions should be considered to justify their interpretation as rest-frame distributions, providing a natural interpolation between the Breit frame and the infinite-momentum frame distributions~\cite{Lor20,Che23}. The latter is consistent with earlier work advocating the same 
correction $3/4M^2$ in the determination of nuclear sizes~\cite{Yen57,Fri97}.

\begin{figure}[h]
   \begin{center}
      \includegraphics[width=1.0\columnwidth]{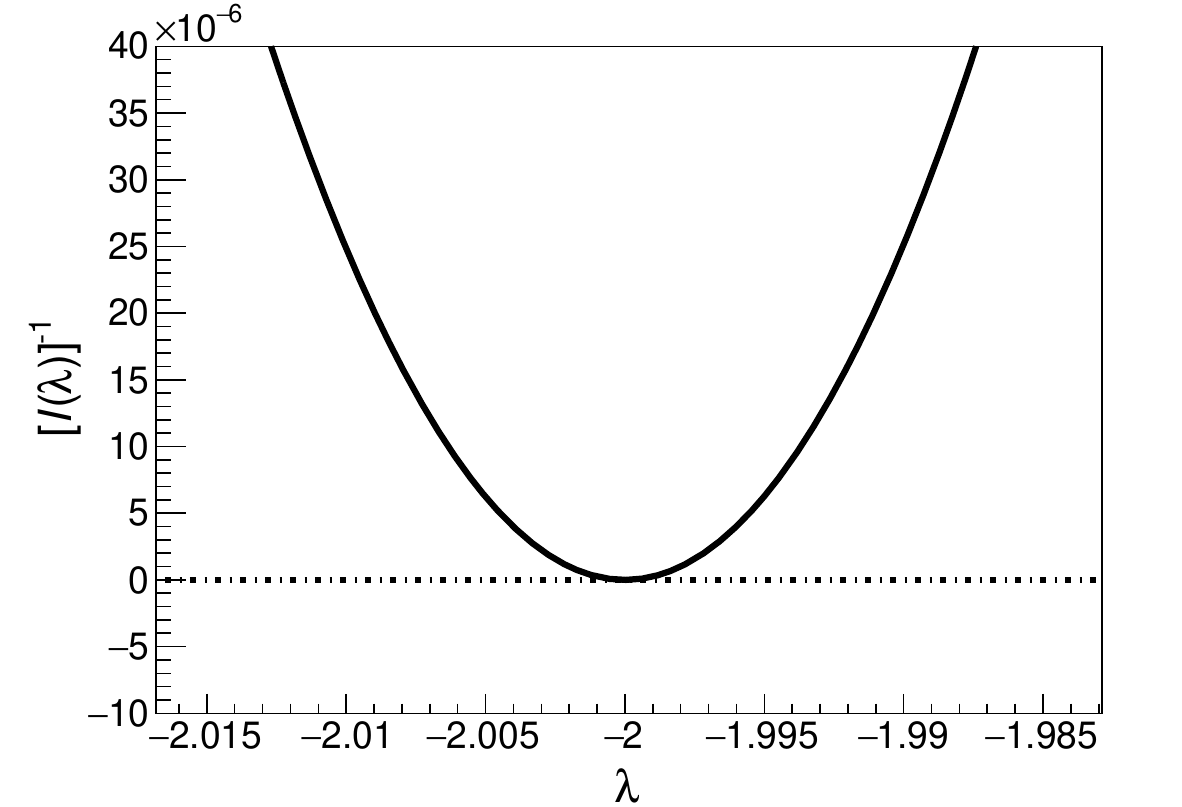}
   \end{center}
\caption{Calculation of the function $\left(\mathcal{I}(\lambda )\right)^{-1}$ for different values of $\lambda$.}
\label{fig:7}
\end{figure}

In this regard, considering cases where the form factor is two-dimensional, we proceed to a case study using the following radial density 
\begin{align}\label{sect4sub4eq01}
f(r) = \frac{a^2}{2\, \pi} K_{0}( a\, r),
\end{align}
satisfying
$2 \, \pi \int_0^\infty f(r) r \, dr = 1$,
where $ K_{0}( a\, r)$ is the modified Bessel Function.
When $ r\to 0$, the series development of this function is given by
\begin{align}\label{sect4sub4eq02}
f(r) \approx
&
\frac{1}{2 \, \pi}
\left( \  -  a ^2  \, \gamma_E -  a ^2  \, \mathrm{\ln}(a/2) \ \right)
- \frac{a^2}{2 \, \pi} \ \mathrm{\ln}(r)
\nonumber \\[1mm]
&
  + 
  \Big(
  \frac{a^4}{8 \, \pi}
  - \frac{a^4 \, \gamma_E}{8 \, \pi} - \frac{a^4 \,\mathrm{\ln}(a/2)}{8 \, \pi}
  \nonumber\\   
  & \hspace{6mm}
  - \frac{a^4}{8 \, \pi} \ \mathrm{\ln}(r)
  \Big) \, r^2 + \cdots,
\end{align}
where $ \gamma_E $ is the Euler-Mascheroni constant.
The Fourier transform is given by 
\begin{align}\label{sect4sub4eq03}
\tilde{f}(k) 
= 2 \, \pi \int_0^\infty f(r) \ r \ J_0( k \, r) \ dr
= \frac{a^2}{a^2 + k^2} ,
\end{align}
and its asymptotic series read
\begin{align}\label{sect4sub4eq04}
\tilde{f}_{\rm asymp}(k)
=
 \sum_{i=i_{\rm min}}^{-2} c_i \ k^{i},
\end{align}
and ($D=2$)
\begin{align}\label{sect4sub4eq05}
 \mathcal{I}(\lambda) =
 \mathcal{N}_{\lambda;2} \
\Bigg\{ 
\int_0 ^{K} 
\frac{\tilde{f}(k)}{k^{ \lambda + 1}}
\ dk
+ 
\sum_{i=i_{\rm min}}^{-2} c_i \ \frac{K^{i - \lambda}}{\lambda  -i}   
 \Bigg\}     .
\end{align}
The value of $ \lambda_c$ is given by  
\begin{align}\label{sect4sub4eq06}
\Big( \mathcal{I}(\lambda_c) \Big)^{-1} = 0
\end{align}
and hence
\begin{align}\label{sect4sub4eq07}
\beta = \lambda_c + 2.
\end{align}
The shape of the function $ \Big( \mathcal{I}(\lambda) \Big)^{-1}$
is shown in Fig.\ref{fig:7} demonstrating a characteristic of a double pole.
For $ \lambda \approx \lambda_c $, we have the following development
(Eq.\ref{sect2sub2eq09}):
\begin{align}\label{sect4sub4eq08}
\mathcal{I}(\lambda) = 
\sum_{j=1}^{n} \frac{B_j}{(\lambda -\lambda_c)^j} .
\end{align}
The value of $ \lambda_c$ being known, we define
the $ \chi^2 $ function 
\begin{align}\label{sect4sub4eq09}
 \chi^2 =
\sum_{i=1}^{2N}
    \Big[ \mathcal{I}(\lambda_i) 
          - f_{\rm th}(\lambda_i,B_{1},B_{2},\cdots, B_{n}) \Big]^2
\end{align}
with
\begin{align}\label{sect4sub4eq10}
f_{\rm th}(\lambda_i,B_{1},B_{2},B_{3},\cdots, B_{n})
\equiv 
 \sum_{j=1}^{n}
 B_{j} \ h(\lambda_i,j)
\end{align}
and
\begin{align}\label{sect4sub4eq11}
h(\lambda_i,j) \equiv \frac{1}{(\lambda_i - \lambda_c)^j}.
\end{align}
The parameters $ B_{1}, B_{2},\cdots, B_{n}$ are determined by minimizing the $  \chi^2 $ function, 
\begin{align}\label{sect4sub4eq12}
\frac{\partial \chi^2 }{\partial B_j} = 0.
\end{align}
We then derive
\begin{align}\label{sect4sub4eq13}
g_{0 \, j-1} = \frac{B_j}{2 \, \pi} \ \frac{(-1)^{j-1}}{(j-1)!}.
\end{align}
The process is used again for $ \lambda \approx \lambda_c -1 $ which gives
\begin{align}\label{sect4sub4eq14}
g_{1 \, j-1} = \frac{B_j}{2 \, \pi} \ \frac{(-1)^{j-1}}{(j-1)!}
\end{align}
and so on. The results of the calculation are given in table \ref{tab:6}.
\begin{table}[h]
\begin{center}
\begin{tabular}{|| c | c  ||} \hline
$ \beta    $ & $ 0.0000 $  \\
$ g_{00}   $ & $ -0.3675$ \\
$ g_{01}   $ & $ -0.6366$ \\
$ g_{10}   $ & $  0.0000$  \\
$ g_{11}   $ & $  0.0000$  \\
$ g_{20}   $ & $  0.2692$  \\
$ g_{21}   $ & $ -0.6366$ \\ 
\hline
\end{tabular}
\caption{Numerical values of $\beta$ and $g_{ij}$'s obtained for $a=2$. They are in full agreement with the analytical ones obtained using Eq.~\ref{sect4sub4eq02}. The value of 
$ \beta = 0$ is equivelant to $\lambda_c = -2$.}\label{tab:6}
\end{center}
\end{table}

We now apply this method to the Dirac form factor.
According to Miller \cite{Miller}, the proton charge density
is given by the Dirac form factor $F_1$ defined as
\begin{align}\label{sect4sub4eq15}
  F_{1}(k^2) = \frac{G_E (k^2) + \tau \, G_M(k^2)}{1 + \tau},
  \end{align}
  with $
\tau = \frac{(\hbar \, c )^2 k^2}{4 \, M^2} $. 
  We notice that
  \ba\label{sect4sub4eq17}
  \lim_{k\to 0}  F_{1}(k^2) 
  &=& \lim_{k\to 0} G_E (k^2), \nonumber
  \\[1mm]
  \lim_{k\to \infty}  F_{1}(k^2) 
  &=&\lim_{k\to \infty} G_M (k^2),
  \ea
  and when $k \to \infty$, we have:
  \be\label{sect4sub4eq18}
  G_E (k^2)
  \ 
  \approx  \sum_{i=i_{\rm min}}^{i=-4}
  C^{e}_{i} \ k^i  
   ,\quad  i  \ \ \mathrm{even},
  \ee
  with  $C^{e}_{-4} = 0$, also when $k \to \infty$:
  \begin{align}\label{sect4sub4eq19}
  G_M (k^2)
  \ 
  \approx 
  \sum_{i=i_{\rm min}}^{i=-4}
  C^{m}_{i} \ k^i   ,\quad  i \ \ \mathrm{even}.
  \end{align}
The inverse Fourier transform gives
\begin{align}\label{sect4sub4eq20}
\mathcal{F}_1 (r) =
\frac{1}{2 \, \pi}
\int_0^{\infty} F_{1}(k^2) \, k \, J_0( kr) \, dk.
\end{align}
If we consider that the contribution of the form factor in momentum space
for large values of $ k^2$ gives the main part of  the radial density 
near $r = 0$, the contribution of the magnetic form factor
to the  radial density near $r = 0$ is important. Keeping this in mind,
we write
\begin{align}\label{sect4sub4eq21}
\mathcal{F}_1 (r) = \mathcal{F}_1^{e} (r) + \mathcal{F}_1^{m} (r)
\end{align}
with
\begin{align}\label{sect4sub4eq22}
\mathcal{F}_1^{e} (r)
\approx \frac{1}{r^{\beta _e}}
\Big[     &      g_{00}^e + g_{01}^e \, \mathrm{\ln}(r) 
      + \left(  g_{10}^e + g_{11}^e \, \mathrm{\ln}(r) \right) \, r
      \nonumber\\
     & + \left(  g_{20}^e + g_{21}^e \, \mathrm{\ln}(r) \right)  \, r^2
      + \cdots
\Big]
\end{align}
and 
\begin{align}\label{sect4sub4eq23}
\mathcal{F}_1^{m} (r)
\approx \frac{1}{r^{\beta _m}}
\Big[      &     g_{00}^m + g_{01}^m \, \mathrm{\ln}(r) 
      + \left(  g_{10}^m + g_{11}^m \, \mathrm{\ln}(r) \right) \, r
      \nonumber\\
     & + \left(  g_{20}^m + g_{21}^m \, \mathrm{\ln}(r) \right)  \, r^2
      + \cdots
\Big].
\end{align}
If $ \beta _e = \beta _m  \equiv \beta$, we have
\begin{align}\label{sect4sub4eq24}
\mathcal{F}_1 (r)
= \frac{1}{r^{\beta}}
\Big[ & g_{00} + g_{01} \, \mathrm{\ln}(r) 
      + \left(  g_{10} + g_{11} \, \mathrm{\ln}(r) \right) \, r 
      \nonumber\\
     & + \left(  g_{20} + g_{21} \, \mathrm{\ln}(r) \right)  \, r^2
      + \cdots
\Big]
\end{align}
with
\[
g_{00} =g^e_{00} + g^m_{00},\;  g_{01} = g^e_{01} + g^m_{01} \cdots, \;
 g_{ij}=g^e_{ij} + g^m_{ij}.
\]
To get the analytical expressions of the  $g_{ij}$'s, we proceed as in the non-relativistic case. The expressions of the electric $G_E (k^2)$ and magnetic form factors $G_M(k^2)$ read:
\begin{align}\label{sect4sub4eq26}
G_E (k^2) &= \frac{1 + \tilde{a}_{1e} \, k^2}
                 {1 + \tilde{b}_{1e} \, k^2 
		    + \tilde{b}_{2e} \, k^4 + \tilde{b}_{3e} \, k^6}
\nonumber \\[1mm]
& = \frac{1}{\tilde{b}_{3e}}
\ \frac{1 + \tilde{a}_{1e} \, k^2}
       {  \prod_{i=1}^3 (k - k_{ie})(k-\bar{k}_{ie})
      }
\end{align}
and
\begin{align}\label{sect4sub4eq27}
G_M (k^2) &= \mu _p \, \frac{1 + \tilde{a}_{1m} \, k^2}
                 {1 + \tilde{b}_{1m} \, k^2 
		    + \tilde{b}_{2m} \, k^4 + \tilde{b}_{3m} \, k^6}
\nonumber \\[1mm]
& = \frac{\mu _p}{\tilde{b}_{3m}}
\ \frac{1 + \tilde{a}_{1m} \, k^2}
       {  \prod_{i=1}^3 (k - k_{im})(k-\bar{k}_{im}) }.
\end{align}
The Partial Fraction Decomposition of $ F_{1}(k^2) \ k $ gives:
\begin{align}\label{sect4sub4eq28}
 F_{1}(k^2) \ k
 = \hspace{4mm}
 &
\sum_{i=1}^4  \left[ \frac{A_{ie}}{k- k_{ie}} + \frac{\bar{A}_{ie}}{k - \bar{k}_{ie}}\right]
\nonumber\\[-1mm]
& \hspace{-3mm}
+ \sum_{i=1}^4  \left[ \frac{A_{im}}{k- k_{im}} 
+ \frac{\bar{A}_{im}}{k - \bar{k}_{im}} \right].
\end{align}
The full calculation of $ \mathcal{F}_1^{e} (r) $ and $ \mathcal{F}_1^{m} (r)$ contains a sum of
modified Bessel functions $K_0$ with complex arguments. Taking their series
for $r \simeq 0$, we obtain
\be\label{sect4sub4eq29}
g_{00}^e
=
- \frac{1}{\pi}
\sum_{j=1}^{n_e}
A_{j e}\ 
\Big( \
 \gamma_E + \mathrm{\ln} \left(- i \, k_{j e}\, /2  \right)
\ \Big)
\ee
with $ n_e =4$ and 
\be\label{sect4sub4eq30}
g_{01}^e
=
- \frac{1}{\pi} \sum_{j=1}^{n_e} A_{j e},
\hspace{12mm}
g_{10}^e = g_{11}^e = 0
\ee
\begin{align}\label{sect4sub4eq31}
g_{20}^e =
\frac{1}{4 \pi}
\sum_{j=1}^{n_e}
 \bigg[
  &
 A_{j e} \, k_{j e}^2 \, 
 \Big( 
  -1 + \gamma_E +  \nonumber\\
  &
  \mathrm{\ln} \big(- i \, k_{j e}/2 \, \big)
  \Big)
 \bigg],
\end{align}
\be\label{sect4sub4eq32}
\text{and} \quad
g_{21}^e =
\frac{1}{4\, \pi}
\sum_{j=1}^{n_e} A_{j e} \, k_{j e}^2 .
\ee
We have  similar relations replacing ${e}$ (electric) by ${m} $ (magnetic).
The roots $ k_{ie}$,  $ k_{im}$ and the coefficients
$A_{ie}$ and  $A_{im}$ have been numerically calculated to determine
the above analytical quantities.
The final results for the coefficients $g_{ij}$'s entering the expressions of radial densities are shown in table \ref{tab:7}. 
We notice that the only coefficient proportional to $ \mathrm{\ln}(r)$ which is not equal to zero is $ g_{21}$ and this term is due to the magnetic form factor.

\begin{table}[h]
\begin{center}
\begin{tabular}{| c  | c | c | c |} 
\hline
$ \ $     & $ \mathcal{F}_1^{e} (r)$        & $ \mathcal{F}_1^{m} (r) $ & $ \mathcal{F}_1 (r) = \mathcal{F}_1^{e} (r) + \mathcal{F}_1^{m} (r) $ \\ 
\hline
$g_{00}$  & $0.9769$    & $ 0.9898 $   &  $ 1.9667  $\\
$g_{01}$  & $0 $        & $ 0   $      &  $ 0  $\\
$g_{10}$  & $0$         & $ 0 $        &  $ 0  $\\
$g_{11}$  & $0$         & $ 0 $        &  $ 0  $\\
$g_{20}$  & $-4.3961$   & $ 6.1234  $  &  $ 1.7273   $\\
$g_{21}$  & $0$         & $ 16.6508$   &  $  16.6508 $\\
\hline 
\end{tabular}
\caption{The numerically evaluated coefficients $g_{ij}$'s corresponding to the electric $ \mathcal{F}_1^{e} (r)$ 
and magnetic $ \mathcal{F}_1^{m} (r)$ radial densities and to the electromagnetic radial density $ \mathcal{F}_1 (r)$ 
corresponding to the Dirac form factor ${F}_1 (k^2)$. These values are in full agreement with the analytically evaluated coefficientss using Eq.~\ref{sect4sub4eq20}.
The values correspond to $ \beta_e = \beta_m = \beta = 0$ which is equivalent to $\lambda_c = -2$.}\label{tab:7}
\end{center}
\end{table}

\section{Conclusions}\label{sect5}

In this work, we have established that, starting from the 
measured form factors, the calculation,
as an integral in momentum space, 
of the moments $(r^\lambda, f_D ) $
with negative value of $\lambda$ gives access to the behavior
of the radial density in the vicinity of $r = 0$.
In a first step, we determine the critical value $\lambda_c$ and
deduce the parameter $\beta$ such as the radial density behaves as $r^{-\beta}$
when $r \to 0$.
Secondly, the method allows the determination of all the terms in the Maclaurin series. This method has been extended to the case 
where logarithms are present in the expansion.
With selected physics cases from the literature, where the series expansions
are analytically known, we have shown that our method is 
numerically accurate.

Experimentally, the determination of negative order moments requires the knowledge of the corresponding form factor at large $k^2$.
In this work we have demonstrated that the higher contribution to the radial density is originating from this domain in $k^2$, putting emphasis
on the importance of directing experiments towards measuring form factors in that domain.

Moreover, our work implies that the chosen functional form of the fit model of the form factor should have the constraint that it should 
vanish in the asymptotic limit; this being a necessary constraint for the existence of the inverse Fourier transform. Nonetheless, even if the
inverse Fourier transform exists, it is not evident that it can be formulated analytically. Of course, one can attempt to extract it numerically 
keeping in mind that Bessel and modified Bessel functions exhibit strong oscillations in the asymptotic limit in $k$ and require in some cases 
the passage to the complex plane.

The radial density being a rapidly decreasing function, we have limited the evaluation, in our examples, to the first, $g_1$ and second, $g_2$ terms of the 
Maclaurin expansion. In such case, it is clear that the knowledge of $g_1$ and $g_2$ - along with $\beta$ - is necessary to know 
the degree of convergence or divergence of the function and whether its concave or convex:
if $g_1>0$ the function is increasing and vice-versa, if $g_2>0$ the function is convex and vice-versa. 
For the case of the modified Bessel function, where the expansion admits logarithmic terms, the knowledge of the expansion 
up to $g_{21}$ is sufficient.

\vspace{20mm}
\begin{acknowledgements}
We are very grateful to Prof. Hagop Sazdjian for fruitful discussions and
encouragements and to Dr. Samuel Friot for the references concerning Mellin
transform.
\end{acknowledgements}

\appendix
\section{Analytical expression of the moments} \label{demo_reg_mom}

In this appendix, we will treat  the regularization of the integral, henceforth denoted as $I$, of the right-hand side of Eq.~\ref{sect3eq05},
\begin{align}\label{sect3eq06}
I =\int_0^{\infty}dk \  \left\{\frac{\tilde{f}(k)}{k^{\lambda+1}}
\right\}_{\infty}.
\end{align}
In the case where $ \lambda > -p$, $I$ needs no regularization at $\infty$ and the integral writes,
\begin{align}\label{sect3eq07}
I &=
\int_0^{K} dk \ \frac{\tilde{f}(k)}{ k^{\lambda+1}}
+ 
\sum_{j=0}^\infty \tilde{f}_{-j}
\int_K^{\infty}  \frac{dk}
                      {k^{\lambda+1 +p +j}}
\nonumber \\
&=
\int_0^{K} dk \ \frac{\tilde{f}(k)}{ k^{\lambda+1}}
+ 
\sum_{j=0}^\infty 
\frac{\tilde{f}_{-j}}{(j+p+\lambda) \, K^{j+p+\lambda}}.
\end{align}
In the case where $ \lambda < -p$, regularizing the integral at $\infty$ is necessary, using Eq.~\ref{sect3eq04} one writes $I$ as
\begin{align}\label{sect3eq08}
I =
   \int_0^{\infty}dk \ 
     \frac{ \tilde{f}(k)-\sum_{j=0}^n \tilde{f}_{-j} \ k^{-j-p}}
          {k^{\lambda+1}}	
\end{align}
where, following the reference \cite{Gue62}, $n$ is the integer part of $-p-\lambda$.
The integral in 
Eq.~\ref{sect3eq08} is decomposed as
\begin{align}\label{sect3eq09}
I &=
 \hspace{4mm}
  \int_0^{K} dk \ 
     \frac{ \tilde{f}(k)-\sum_{j=0}^n \tilde{f}_{-j} \ k^{-j-p}}
        {k^{\lambda+1}}
\nonumber \\
& \hspace{3mm}
 +  \int_K^{\infty} dk \
     \frac{\sum_{j=n+1}^\infty \tilde{f}_{-j} \ k^{-j-p}}
          {k^{\lambda+1}}
\nonumber \\	  
&	  
=
 \hspace{4mm}
   \int_0^{K} dk \ 
     \frac{ \tilde{f}(k)-\sum_{j=0}^n \tilde{f}_{-j} \ k^{-j-p}}
          {k^{\lambda+1}}
\nonumber \\
& \hspace{3mm}
+
\sum_{j=n+1}^\infty
      \frac{ \tilde{f}_{-j} } 
           {(\lambda + p + j) \, K^{\lambda + p + j}}	  	.  	
\end{align}
If we consider the case $ -p-1 < \lambda < -p$. We then have $n=0$, and
\begin{align}\label{sect3eq10}
I 
= 
&
\int_0^K dk \frac{\tilde{f}(k)-\tilde{f}_{0}\,  k^{-p}}{k^{\lambda+1}}
\nonumber \\
& \hspace{2mm}
 + 
\sum_{j=1}^\infty
      \frac{ \tilde{f}_{-j} }
           {(\lambda + p + j) \, K^{\lambda + p + j}}.
\end{align}
We notice that 
\begin{align}\label{sect3eq11}
- \tilde{f}_{0}
\ \int_0^K \frac{dk}{k^{\lambda+1+p}}
= 
\frac{ \tilde{f}_{0} }
           {(\lambda + p ) \, K^{\lambda + p }},
\hspace{6mm} \lambda < -p
\end{align}
and Eq.~\ref{sect3eq10} reads
\begin{align}\label{sect3eq12}
I 
= 
&
\int_0^K dk \frac{\tilde{f}(k)}{k^{\lambda+1}}
 + 
\sum_{j=0}^\infty
      \frac{ \tilde{f}_{-j} }
           {(\lambda + p + j) \, K^{\lambda + p + j}},
\end{align}
which is exactly the same as Eq.~\ref{sect3eq07}
We now consider the complementary case $ -p-2 < \lambda < -p - 1$ where we then have $n=1$ and
\begin{align}\label{sect3eq13}
I 
= 
&
\int_0^K dk \frac{\tilde{f}(k)-\tilde{f}_{0}\,  k^{-p} 
                  -\tilde{f}_{-1}\,  k^{-p-1}   }
                 {k^{\lambda+1}}
\nonumber \\
& \hspace{2mm}
 + 
\sum_{j=2}^\infty
      \frac{ \tilde{f}_{-j} }
           {(\lambda + p + j) \, K^{\lambda + p + j}}.
\end{align}
We notice that 
\begin{align}\label{sect3eq14}
&
- \tilde{f}_{0}  \ \int_0^K \frac{dk}{k^{\lambda+1+p}}
- \tilde{f}_{-1} \ \int_0^K \frac{dk}{k^{\lambda+2+p}}
\nonumber \\
&
= 
\frac{ \tilde{f}_{0} }
     {(\lambda + p ) \, K^{\lambda + p }}
+	   
\frac{ \tilde{f}_{-1} }
     {(\lambda + p +1 ) \, K^{\lambda + p +1}},	   	   
\hspace{6mm} \lambda < -p -1
\end{align}
and Eq.~\ref{sect3eq13}
becomes
\begin{align}\label{sect3eq15}
I 
= 
&
\int_0^K dk \frac{\tilde{f}(k)}{k^{\lambda+1}}
 + 
\sum_{j=0}^\infty
      \frac{ \tilde{f}_{-j} }
           {(\lambda + p + j) \, K^{\lambda + p + j}},
\end{align}
which is exactly the same as Eq.~\ref{sect3eq07} and Eq.~\ref{sect3eq12}. Finally, spatial moments are expressed as:
\begin{align}\label{sect3eq16}
 (r^{\lambda},f_D) 
 &=
 2^{\lambda+1}
 \frac{\Gamma \left(\frac{\lambda+D}{2} \right)}
      {\Gamma \left( -\frac{\lambda}{2} \right) 
        \Gamma \left( \frac{D}{2}\right)}
\int_{0} ^{\infty} dk \, { \left\{ \frac{ \tilde{f}(k) }
                                        {k^{\, \lambda +1 }} 
		           \right\} }_{\infty}
 \nonumber \\
& 
=
 2^{\lambda+1}
 \frac{\Gamma \left(\frac{\lambda+D}{2} \right)}
      {\Gamma \left( -\frac{\lambda}{2} \right) 
        \Gamma \left( \frac{D}{2}\right)}
\nonumber \\
& \hspace{-10mm}
  \times
\Bigg[
  \int_0^{K} \! \! \! dk 
  \,  \frac{\tilde{f}(k)}{k^{\lambda+1}} 
  + \sum_{j=0}^{\infty} \frac{\tilde{f}_{-j}}
                       { \, (  \lambda+p+j) \, K^{\lambda+p+j}}
\Bigg].
\end{align}



\end{document}